\documentclass[11pt]{article}
\usepackage{geometry}                
\usepackage{graphicx}
\usepackage{amssymb}
\usepackage{epstopdf}
\def \lsim{\mathrel{\vcenter
     {\hbox{$<$}\nointerlineskip\hbox{$\sim$}}}}
\def \gsim{\mathrel{\vcenter
     {\hbox{$>$}\nointerlineskip\hbox{$\sim$}}}}

\begin{document}
\thispagestyle{empty}

\vspace{40pt}
\begin{center}

\textbf{\Large Signatures of TeV gravity from the evaporation
 of cosmogenic black holes\footnote{Talk presented by I.M. at the Workshop on Black Holes in General Relativity and String Theory,
 August 24-30 2008, Veli Lo\v{s}inj, Croatia.}}
\vspace{40pt}

Iacopo Mastromatteo$^{a}$, Petros Draggiotis$^{b}$, Manuel Masip$^{b}$

\vspace{12pt}

\textit{
$^{a}$Dipartimento di Fisica Teorica}\\ 
\textit{Universit\`a degli Studi di Trieste, I-34014 Trieste, Italy}\\
\vspace{8pt}
\textit{$^{b}$CAFPE and Departamento de F{\'\i}sica Te\'orica y del
Cosmos}\\ 
\textit{Universidad de Granada, E-18071 Granada, Spain}\\
\vspace{16pt}
\texttt{mastroma@sissa.it, pdrangiotis@ugr.es, masip@ugr.es}
\end{center}

\vspace{40pt}

\date{\today}

\begin{abstract}

TeV gravity models provide a scenario for black 
hole formation at energies much smaller than 
$G_N^{-1/2} \sim 10^{19}$ GeV.  In particular, the collision of 
a ultrahigh energy cosmic ray with a dark matter particle in our 
galactic halo or with another cosmic ray could result into a 
black hole of mass between $10^4$ and $10^{11}$ GeV. 
Once produced, such object would evaporate into elementary
particles via Hawking radiation. We show that the 
interactions among the particles exiting the black hole 
are not able to produce a photosphere nor a chromosphere.
We then evaluate how these particles evolve using the jet-code
HERWIG, and obtain a final 
diffuse flux of stable 4-dimensional particles peaked at 
0.2 GeV. This flux consists of an approximate $43\%$ of neutrinos, 
a $28\%$ of electrons, a $16\%$ of photons and a $13\%$ of 
protons. Emission into the bulk would range from a $1.4\%$ 
of the total energy for $n=2$ to a $16\%$ for $n=6$.

\end{abstract}

\newpage

\section{Introduction}
Models with extra dimensions \cite{ArkaniHamed:1998rs} 
provide one of the most promising 
solutions to the hierarchy problem, namely, the huge difference 
between the scale of gravity $M_P = G_N^{-1} \sim 10^{19}$ GeV 
and the electroweak (EW) scale $M_{EW} \sim 100$ GeV. In these 
models $M_P$ appears as an effective scale related
with the fundamental one, $M_D \sim$ 1--10 TeV, by the 
volume of the compact space or by an exponential warp factor. 
The difference between $M_{EW}$ and $M_D$ would then just define
a {\it little} hierarchy problem that should be easier to
solve consistenly with all collider data.
The phenomenological consequencies of this framework are quite
{\it intriguing}:
the fundamental scale would be at accessible 
energies, and processes with 
$\sqrt{s} \gg M_D$ would probe a {\it transplanckian} regime  
where gravity is expected to dominate over the other 
interactions \cite{Emparan:2001ce}. 
The spin two of the graviton implies then gravitational cross sections that 
grow fast with $\sqrt{s}$ and become long distance
interactions. As a consequence, quantum gravity or other 
short distance effects become irrelevant as they are screened by 
black hole (BH) horizons \cite{Giddings:2001bu}.

One of the scenarios in which TeV gravity effects could  
play a significant role is provided by cosmic rays physics. The Earth 
is constantly hit by a flux of protons with energy of up to 
$10^{11}$ GeV and, associated to that flux, it is also expected a
flux of cosmogenic neutrinos (still unobserved) with 
a typical energy peaked around $10^{10}$ GeV \cite{Semikoz:2003wv}.
These are energies much larger than the ones to be explored
at the LHC, where there would be no evidence for gravitational 
interactions if the scale $M_D$ is above a few TeV.
In addition, notice that the new physics should be more relevant in
collisions of particles with a small SM cross section, as
it is expected for the interaction of a proton with a dark matter 
particle $\chi$ if it is taken to be a weakly interacting massive 
particle (WIMP). 
We will discuss here the interaction of ultra high energy cosmic 
rays (UHECR) with dark matter particles $\chi$ in our galactic
halo. No detail about the nature of $\chi$ other than its mass,
which defines the center-of-mass energy 
$\sqrt{s}=\sqrt{2m_\chi E}$ in the collision, 
is going to be significant to the present analysis.
We also consider collisions of UHECR with 
other cosmic rays. These are arguably the most energetic 
elementary processes that we know that occur in nature
at the present time, and would produce mini BHs significantly
colder and longer-lived than the ones usually considered in
the literature.
We will focus just on BH production and evaporation, being
this analysis a necessary first step in order to 
understand the full effects of TeV gravity on UHECR phenomenology.

\section{Cosmogenic black hole production}
\begin{figure}
\begin{center}
\includegraphics[width=0.5\linewidth]{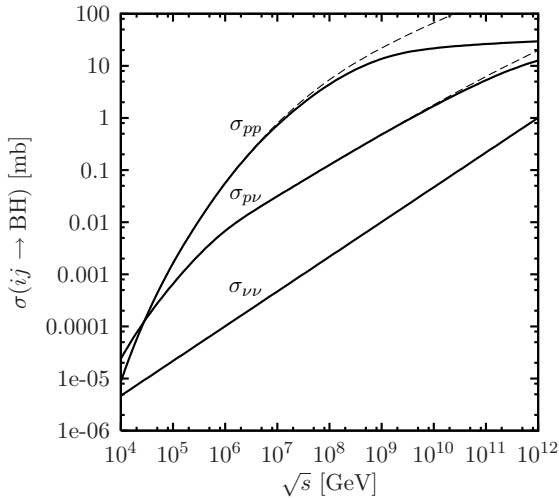}
\caption{
Cross sections to produce a BH for $n=2$ and $M_D=1$ TeV. 
\label{fig1}}
\end{center}
\end{figure}

BH production processes are the most widely and 
detailfully discussed aspect of TeV-gravity phenomenology 
\cite{D'Eath:1992hb}, and they have been considered both in the LHC 
\cite{Dimopoulos:2001en} and in the UHECR context \cite{Feng:2001ib}. 
Here we will assume a scenario with $n$ flat extra dimensions of common 
lenght where gravity is free to propagate, while matter fields are 
trapped on a (non-compact) four-dimensional brane. 
We will use the basic estimate 
that the collision of two pointlike particles at impact 
parameters smaller than the Schwarzschild radius $r_H$ 
of the system leads 
to the production of a BH whose mass is given by $M=\sqrt{s}$. 
The BHs that we are considering 
($M < 10^{11}$ GeV) will be described by a ($4+n$)-dimensional 
metric (they are smaller than the volume of the compact space), 
being their radius
\begin{equation}
r_{H}=\left({2^n\pi^{n-3\over 2}\Gamma\left({n+3\over 2}\right)
\over n+2}\right)^{1\over n+1}
\left({M\over M_D}\right)^{1\over n+1}
     {1\over M_D}\;.
\label{rh}
\end{equation}
For two pointlike particles, the cross section 
$\sigma (s)= \sigma_{\nu \nu}= \sigma_{\nu \chi}$ 
to produce a BH is then written as
\begin{equation}
\sigma = \pi \, r_H^2 \; .
\end{equation}
If the collision involves non-elementary (at the scale $\mu=1/r_H$)
protons, then its partonic structure has to be included in order to 
find the total cross section, as it is usually done for analyses of
BH production at LHC \cite{Dimopoulos:2001en}.
The p--$\chi$ (or p--$\nu$) cross section   
may therefore be written as
\begin{equation}
\sigma_{p\nu}(s)=\int_{M_D^2/s}^1 {\rm d}x 
\left(\sum_i f_i(x,\mu) \right) \hat \sigma(xs)\;. 
\label{sigmapnu}
\end{equation}
This formula expresses the cross section as the sum of 
partial contributions $\hat \sigma(xs)$ to produce a BH of mass 
$M= \sqrt{xs}$ resulting from the collision of a parton 
$i$ that carries a fraction $x$ of momentum with a pointlike target. 
It is crucial to notice that the scale $\mu$ in the collision
is fixed by the inverse Scwarzschild radius, rather than by the 
BH mass \cite{Giddings:2001bu} \cite{Emparan:2001kf}, since the 
scattering is probing a 
lenght scale that grows (not decreases!) with $s$. Actually, we 
expect that for large enough $s$ the scale that we are exploring goes 
above its radius and a pointlike behaviour for the proton 
will emerge. In contrast with a QED scattering, here at lower 
energies 
($\approx 10^3$ GeV) we can {\it see} the composite structure 
of the proton, while at higher energies ($\approx 10^9$ GeV) 
the proton will scatter coherently as a whole. 
Since Eq. \ref{sigmapnu} does not reproduce this behaviour, 
it is necessary to include matching corrections between the two
energy regions. The cross section in Eq. \ref{sigmapnu} describes
the low-energy regime, and it is dominated by the large number of 
partons of low $x$ that may produce a BH of mass near the threshold 
$M_D$. This scheme explains why  
$\sigma_{p\nu} > \sigma_{\nu\nu}$.
When the cross section $\sigma_{p \chi}$ approaches the proton size
($\approx 20$ mbarn), then the
density of partons with enough energy to produce a BH is so large
that the parton cross sections overlap, and 
the BHs produced are big enough to trap 
other {\it spectator} partons. This overlapping reduces
the total cross section and increases the average 
mass of the produced BH. In this regime 
$\sigma_{p\nu}$ is basically constant with $s$ until it matches  
the pointlike behaviour in $\sigma_{\nu\nu}$.
A similar behavior is also expected in $p$--$p$ collisions, 
where the partonic 
enhancement of the cross section is even more important 
at lower energies (in this regime 
$\sigma_{pp} > \sigma_{p\nu} > \sigma_{\nu\nu}$) and the 
intermadiate regime of constant total cross section is reached 
at lower energies.
The smooth transition from these regimes can be 
modelled numerically discounting the 
contributions from spectator partons, and are 
summarized in Fig.~\ref{fig1}. There we plot
the BH production cross section for different kind of 
particles\footnote{We assumed a CTEQ6M 
set of PDF \cite{Pumplin:2002vw}}.

We will analyze two processes that can lead to BH production
(see \cite{Draggiotis:2008jz} for the fluxes of proton, 
cosmogenic neutrinos and for the dark matter density).

{\it (i)} A cosmic ray of energy $E$  colliding with a dark 
matter particle 
$\chi$ at rest in the frame of reference 
of our galaxy. The average number of BHs produced per unit time 
and volume depends on the density $\rho_\chi$, 
the cross section $\sigma_{i\chi} $ and the differential flux of 
cosmic rays ${\frac{d\phi_i}{dE}}$ (with $i={\rm p},\nu$):
\begin{equation}
{{\rm d}^2N\over {\rm d}t\; {\rm d} V}= 
4 \pi \int {\rm d}E \;\sigma_{i \chi} (s)\;
\frac{{\rm d}\phi_i}{{\rm d}E}\; \rho_\chi \;.
\label{eqDMCR}
\end{equation}
Here the center of mass energy 
$\sqrt{s}=\sqrt{2 m_\chi E}$ can
run from $M_D$ to $10^7$ GeV.

{\it (ii)} A cosmic ray of energy $E_1$ colliding with 
a cosmic ray of energy $E_2$. In this case the center of mass 
energy depends upon the relative angle $\theta$, and results into
$\sqrt{s}=\sqrt{2 E_1 E_2(1-\cos{\theta})}$. The interaction rate 
per unit time and volume is expressed by:
\begin{equation}
{{\rm d}^2N\over {\rm d}t\; {\rm d} V}= 16 \pi^2 
\int {\rm d}E_1\; {\rm d}E_2\; {\rm d}\cos\theta
\;\sigma_{ij} (s)\;\sin\theta/2\;
\frac{{\rm d}\phi_i}{{\rm d}E_1} \;
\frac{{\rm d}\phi_j}{{\rm d}E_2}\;.
\label{eqCRCR}
\end{equation}
These processes generate BH masses  
$M=\sqrt{s}$ that can reach $\sim 10^{12}$ GeV.

\begin{figure}
\begin{center}
\includegraphics[width=0.5\linewidth]{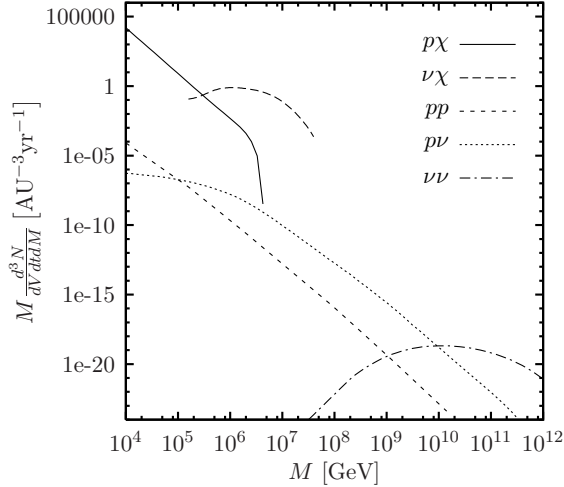}
\caption{
Spectrum of BHs produced by collisions of cosmic rays (protons
and cosmogenic neutrinos) when $n=2$, $M_D=1$ TeV, $m_\chi
=100$ GeV . 
\label{fig2}}
\end{center}
\end{figure}

In Fig.~\ref{fig2} we plot the production rate of BHs from
both types of collisions.

\section{Black hole evaporation}
To understand what kind of signal one could observe from
such an event, it is necessary to estimate how the BH evolves
after its production. 
It is expected that initially the BH undergoes a quick  
{\it balding} phase, in which it loses its gauge hair and  
asymmetries. Then it experiences a {\it spin down} phase, where 
its angular momentum is radiated while losing just a small fraction of 
its mass \cite{Winstanley:2007hj}. Finally, during most of its 
life the BH is in a Schwarzschild phase, losing mass through 
spherically symmetric Hawking radiation \cite{Hawking:1974rv}.
The spectrum is, in a first approximation, that of a black body of 
temperature \cite{Myers:1986un}
\begin{equation}
T={n+1\over 4\pi r_{H}}\;.
\label{eqT}
\end{equation}
This means that the scale of emission is fixed by the inverse 
Schwarzshild 
radius. This formula has important corrections arising 
from the gravitational barrier that the
particles have to cross once emitted. These corrections are usually
expressed in terms  
of the so called {\it greybody factors}, effective emission areas 
$\sigma_n^{(i)} (\omega)$ that depend on 
the dimensionality $(4+n)$ of the space-time, the spin of the
particle emitted, and its energy $\omega$ \cite{Page:1976df}. These
factors give corrections of order 1 to the black-body emission rates 
for all particles species except for the graviton, 
which can have a stronger correction depending upon the number of 
extra dimensions. We will 
assume here the numerical greybody factors given in \cite{Harris:2003eg}.

The number of particles of the 
species $i$ emitted with ($4+n$)-dimensional momenta between $k$ 
and $k+{\rm d}k$ in a time interval ${\rm d} t$ can be written as
\begin{equation}
{\rm d}N_i(\omega) = g_i \, \sigma_n^{(i)}(\omega) \left( \frac{1} 
{\exp{(\omega/T_{BH})} \pm 1} \right) \frac{{\rm d}^{n+3} k}{(2\pi)^{n+3}} \,
{\rm d}t\;,
\label{eqnumber}
\end{equation}
while the radiated energy is given by
\begin{equation}
{\rm d}E_i(\omega) = g_i \, \sigma_n^{(i)}(\omega) \left( \frac{\omega} 
{\exp{(\omega/T_{BH})} \pm 1} \right) \frac{{\rm d}^{n+3} k}{(2\pi)^{n+3}} \,
{\rm d}t\;.
\label{eqenergy}
\end{equation}
Some remarks are here in order.

{\it (i)} On dimensional grounds $\dot{E}\sim A_{2+n} 
T^{4+n}\sim 1/r_{H}^2\sim T^2$ and $\dot{N} \sim T$, so each 
degree of freedom should contribute equally (up to order
one geometric and greybody factors) to the total 
emission independently from its bulk or brane 
localization \cite{Emparan:2000rs}. 

{\it (ii)} We are considering BH temperatures above 
$\Lambda_{QCD}$ ($M \lsim 10^{11}$ GeV leads to $T \gsim 1$ GeV), 
so QCD degrees of freedom (quarks and gluons) are also radiated 
and dominate the total emission.

\begin{figure}
\begin{center}
\includegraphics[width=0.5\linewidth]{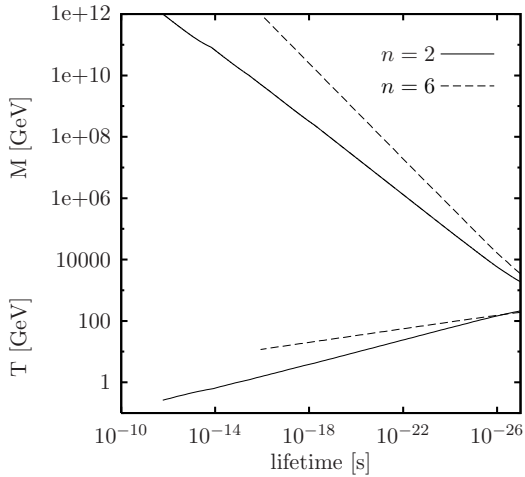}
\caption{
Correlation between mass, temperature,
and lifetime of a BH for $M_D=1$ TeV and $n=2,6$. 
\label{fig3}}
\end{center}
\end{figure}

Once the instant spectrum is known, we integrate it over 
time to get the BH lifetime. On dimensional grounds 
$\tau \sim M_D^{-1} \left( M/M_D \right)^{\frac{n+3}{n+1}}$, 
although the dependence upon the number of the radiated degrees of 
freedom at different temperatures may be significant. In Fig.~\ref{fig3}
we plot the correlation between lifetime, mass and initial 
temperatures for BHs of mass ranging from $10$ TeV to $10^{11}$ GeV, 
$n=2,6$ and $M_D=1$ TeV; it is there shown that lifetimes go 
from a maximum of $10^{-14}$ s for the most heavy BHs to a 
minimum around $10^{-26}$ s for LHC-like BHs.

\section{Thermal properties of the radiation}
An important issue about the evolution of the radiation from
a BH is the debated question relative to its thermalization. It 
has been argued \cite{Heckler:1995qq} that the emitted particles 
should produce a thick shell of almost-thermal plasma of QED 
(QCD) particles usually called photosphere (chromosphere). This 
would occur for BHs above a critical temperature 
$T_{QED}$ ($T_{QCD}$), and would change the average 
energy of the emitted particles from $E_{av} \sim T$ to 
$E_{av} \approx m_e$ (or $E_{av} \approx \Lambda_{QCD}$). 
The argument leading to these shells is 
based on the average number $\Gamma $ of interactions of the 
particles exiting the BH, so $\Gamma\gg 1$ should suffice
to confirm the presence of the plasma shell. 
Initial estimates \cite{Heckler:1995qq} used the expression 
\begin{equation}
\Gamma = \langle \sigma v \rho \rangle \,
\label{gamma}
\end{equation}
which describes the case of particles scattering against a fixed target. 
Recently, however, it has been noticed that the kinematic
differences between that case and the case of particles 
exiting radially from a BH are so significant that lead to a 
complete suppression of the interaction rate \cite{MacGibbon:2007yq}.
We will show, following the approach of Carr, McGibbon and Page, 
that their arguments\footnote{These arguments are supported
by the numerical analysis in \cite{Alig:2006up}.} (formulated 
for ordinary BHs) hold also for BHs in TeV gravity models. 

The first kinematic effect is due to {\it causality}, 
and depends on the fact that the scattered particles
do not come from infinity (as in a regular collision), they
are created in definite points of space-time. This 
introduces a {\it minimal} separation between particles 
successively emitted, both in time and lenght, and induces via 
Heisenberg's indetermination principle an UV cutoff in the 
scale of the exchanged momenta. The scattering cross section
is reduced because not all the energies can be  
interchanged. In particular, in QED (QCD)  
Bremsstrahlung and pair production the momenta dominating 
the collision are of order 
$Q^2 \sim m_e^2$ (or right above $\sim \Lambda_{QCD}^2$). 
If the particle wave functions do not overlap, and their
minimum distance $\nu^{-1}$ (in units of their Compton wavelength) 
is larger than the dominant inverse momenta,
then the process will be suppressed.
Checking the parameter $\nu$ is sufficient to decide about 
the effective connection between emitted particles, and eventually 
exclude thermal interactions. In \cite{Draggiotis:2008jz}
We have shown that this argument excludes the presence 
of a photosphere for any number of extra dimensions, but not 
of a chromosphere when $n>2$.

The second suppression effect is based on the fact that the 
interaction between two particles is not instantaneous, it 
takes a finite time to complete. It is easy to see that when
this occurs the particles are already far away from each other,
so that they can not interact again.
In particular, after completing a QCD interaction partons will 
be at distances larger than $\Lambda_{QCD}^{-1}$, where QCD is
already ineffective. To fully understand 
this point, one first has to notice that the interaction 
between particles moving radially in the same direction
(within the {\it exclusion cone}) is negligible, 
as the density in such a region is low. Also, that
particles moving radially keep on moving radially, 
as the average angular deviation due to Bremsstrahlung-like 
processes is small. This implies that the distance of a
particle to the particles out of the exclusion cone
will always increase (they {\it never} approach to each other), 
and when it reaches a radius $r_{brem}$ this distance 
will be bigger than $m_e^{-1}$ (or $\Lambda^{-1}_{QCD}$)
and the particle is no longer able to interact. 
If the BH temperature is above $T \sim \Lambda_{QCD}$, 
as it is the case for the BHs under study here, it is easy 
to see that after the particle has completed one interaction it
will have already crossed $r_{brem}$.

\section{Stable particle spectrum}
Once the greybody spectrum of emission has been established, 
it is necessary to study how it evolves at astrophysical distances:
unstable particles will decay, and colored particles (which dominate
the spectrum) will fragment into hadrons and then shower into 
stable species. We present our results following 
the approach of \cite{MacGibbon:1990zk}, who first studied 
this issue for primordial BHs.
The main difference with their analysis is that while 
the authors in 
\cite{MacGibbon:1990zk} compute the stationary spectrum at a 
given $T$ (which only changes on astrophysical time scales), we 
need here to evaluate the spectrum integrated over 
the whole (very short) BH lifetime. In any case, our results 
will be analogous,  
since the temperature of a BH variates little for most of its 
lifetime. Of course, our framework also deals 
with a different scale of gravity $M_D \ll M_{Pl}$ and extra
dimensions where gravitons propagate. This implies emission into
the bulk and different greybody factors for all the species. Notice
finally that the spectrum that we are discussing is in the BH rest
frame, it is {\it not} the one to be observed at the Earth as 
the BHs produced in cosmic ray collisions will be highly boosted. 

\begin{figure}
\begin{center}
\begin{tabular}{cc}
\includegraphics[width=0.5\linewidth]{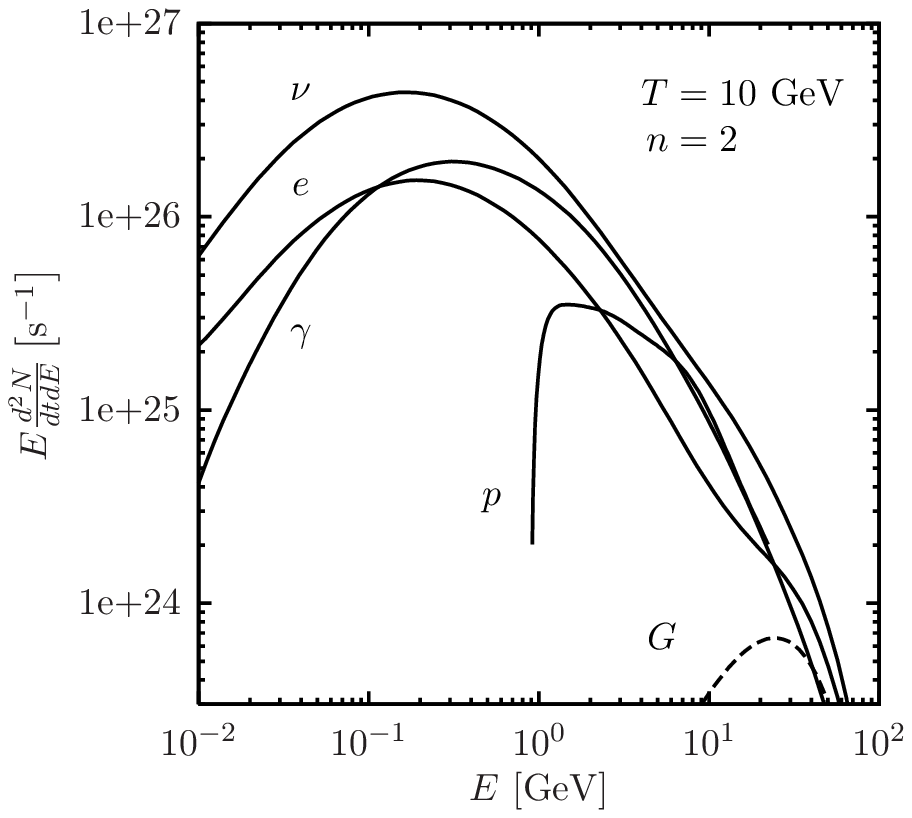} &
\includegraphics[width=0.5\linewidth]{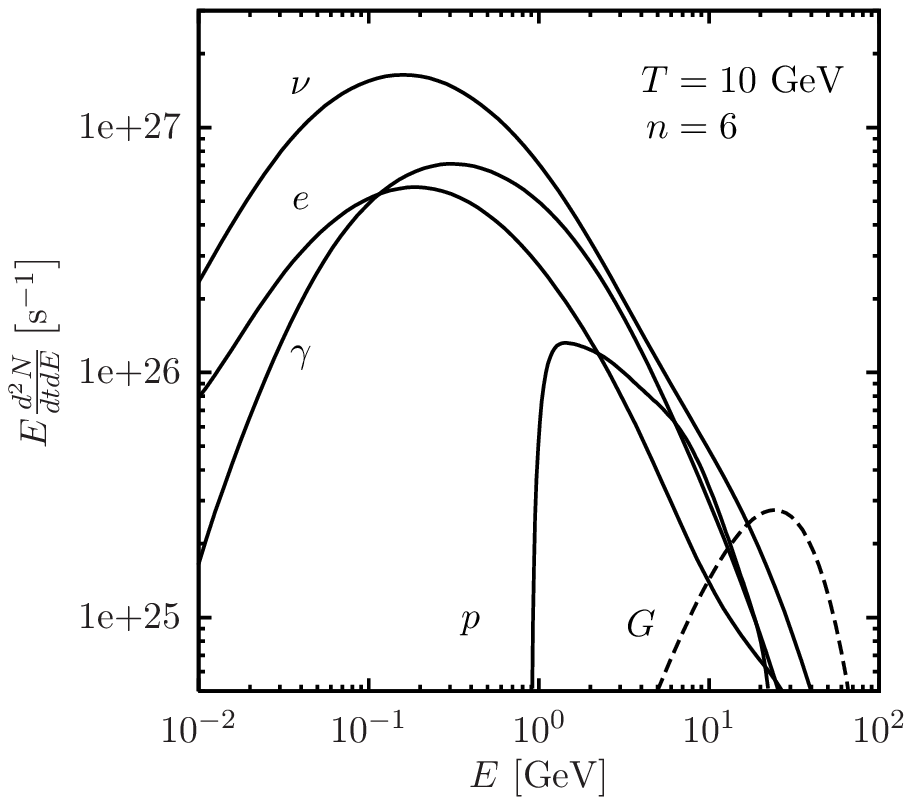} 
\end{tabular}
\end{center}
\caption{Instant spectrum of stable particles and bulk gravitons (dashed) 
emitted by a BH of temperature $T=10$ GeV for $M_D=1$ TeV 
and $n=2$ (left) and $n=6$ (right).
\label{fig4}}
\end{figure}

\begin{figure}
\begin{center}
\begin{tabular}{cc}
\includegraphics[width=0.5\linewidth]{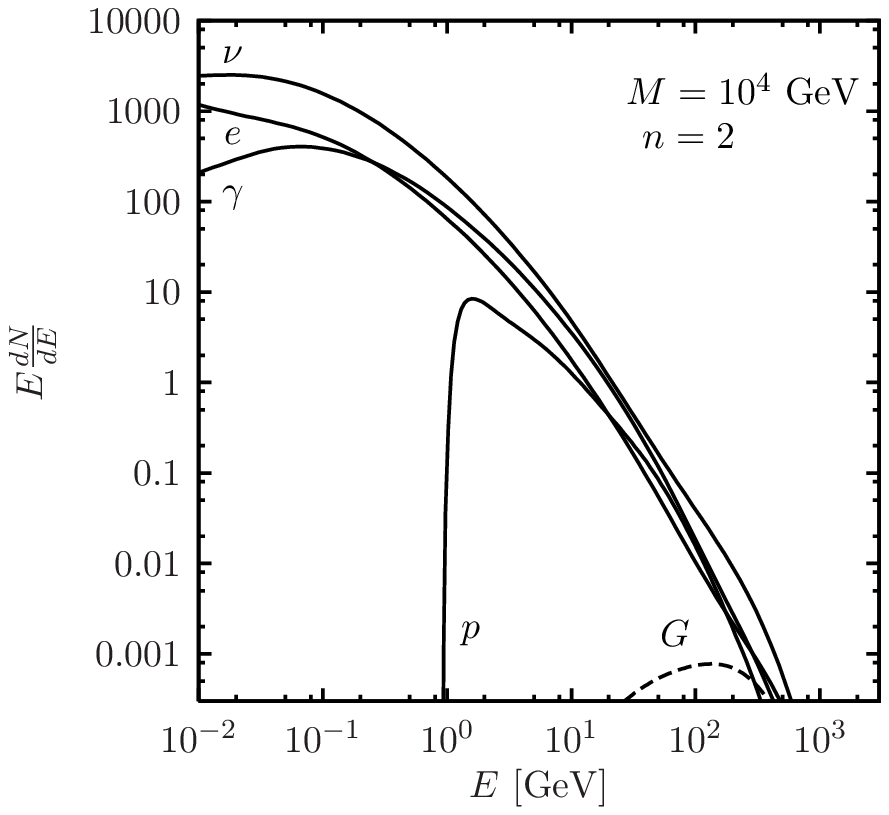} &
\includegraphics[width=0.5\linewidth]{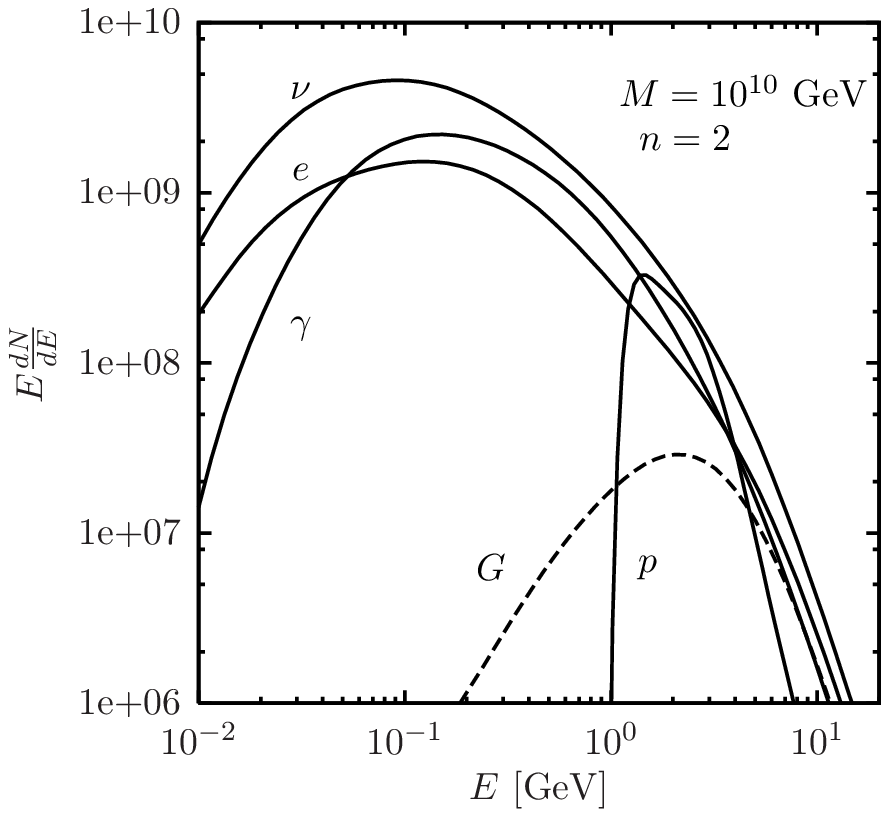} 
\end{tabular}
\end{center}
\caption{Total spectrum of stable particles and bulk gravitons (dashed) 
produced by a BH of mass $M=10$ TeV (left) and $M=10^{10}$ GeV 
(right) for $M_D=1$ TeV and $n=2$.
\label{fig5}}
\end{figure}

We will assume that the evolution of the 
species $i$ emitted by a BH at rest coincides with the one 
in $e^+ e^- \rightarrow i \bar i$ in the center of mass frame, 
so we will use the MonteCarlo jet code HERWIG6 \cite{Corcella:2000bw} 
to evolve the greybody spectrum described before. 
Namely, we compute the convolution
 \begin{equation}
 \frac{{\rm d}N_j}{{\rm d}t {\rm d}\omega} = 
\sum_i \int {\rm d}\omega^\prime \left( 
\frac{{\rm d}N_i}{{\rm d}t {\rm d}\omega^\prime} 
(\omega^\prime)\right) \, 
\left( \frac{{\rm d} g_{ji}}{{\rm d}\omega} 
(\omega, \omega^\prime) \right) \,,
 \end{equation}
to obtain the number ${\rm d}N_j$ of stable particles of species $j$ with 
energy between $\omega$ and $\omega+{\rm d}\omega$ emitted in a time 
${\rm d}t$. The first term in parenthesis stands for the greybody 
spectrum of emission for particle species $i$, while the second 
encodes the probability for the species $i$ of energy $\omega^\prime$ 
to give a $j$ of energy $\omega$. For a given $T$, this has been 
implemented via MonteCarlo including all 
particles of mass $m_i < T$ (leptons, quarks and 
gauge bosons, neglecting the Higgs or the dark matter particle) and 
has resulted in a final spectrum of neutrinos, electrons, photons and 
protons. The spectrum includes the same number of particles and 
antiparticles (they are generated at the same rate), and the three 
families of neutrinos (their flavor oscillates at astrophysical 
lenght scales).
In Fig.~\ref{fig4} we plot the spectrum at fixed temperature $T=10$ GeV,
whereas in Fig.~\ref{fig5} we give the complete spectrum for initial 
masses of $M=10^4$ GeV and $10^{10}$ and $n=2$.
The results 
can be summarized as follows.

{\it (i)} The main product of the emission is constituted 
by particles resulting from the showering of QCD species; this 
explains the primary peak at $\approx 0.2$ GeV observed in 
the spectrum. It is also possible to detect at $E \sim T$ 
the direct greybody emission as a secondary peak. Gravitons 
{\it decouple}, since they are not produced by decay of 
unstable species.

{\it (ii)} The relative emissivities of Standard Model 
particles are an approximate $43\%$ of neutrinos, a $28\%$ 
of electrons, a $16\%$ of photons and a $13\%$ of protons. This is 
only mildly sensitive to the BH mass or $M_D$, as it is 
determined by the showering of colored particles.

{\it (iii)} Emission into the bulk goes from the 0.4\% of 
the total number of particles (16\% of the total energy) 
emitted if $n=6$ and $T= 1.2$ GeV to the 0.02\% of the particles 
(1.4\% of the energy) emitted for $n=2$  and $T=120$ GeV.

\section{Summary and outlook}
The head to head collision of two cosmic rays provides center-of-mass
energies of up to $10^{11}$ GeV. In models with extra dimensions
and a fundamental scale of gravity at the TeV such collision should
result in the formation of a mini BH. Its evaporation and showering 
into stable particles could provide an observable signal.

We have estimated the production rate of these BHs (Fig.~\ref{fig2}) via 
collisions of two cosmic rays or, more frequently, in the
collision of a cosmic ray and
a dark matter particle. In particular, it seems worth to analyze the
possibility that 
{\it (i)} extragalactic cosmic rays crossing the galactic DM halo 
produce a flux of secondary particles with a characteristic shape 
and strongly dependent upon galactic latitude; {\it (ii)} a fraction 
of the flux of cosmic rays with energy up to $\sim 10^8$ GeV trapped 
in our galaxy by $\mu G$ magnetic fields  can be processed by TeV 
interactions into a secondary flux peaked at smaller energies. 
Notice that the physics proposed in this talk is expected 
to become relevant just at center of mass energies above 
$\sqrt{s}\sim \sqrt{2 E m_\chi} \sim 1$ TeV, {i.e.}, at 
cosmic ray energies around the cosmic ray {\it knee}. 
These considerations will be worked out in \cite{Masip:2008mk}, where the 
additional effects of 
gravitational elastic interactions will also be included.

Here we have discussed the
properties of BHs with masses between $10^4$ and $10^{11}$ GeV. 
Such objects have a proper lifetime
between $10^{-14}$ and $10^{-26}$ s (Fig.~\ref{fig3}), and their 
desintegration 
products are mainly determined by the fragmentation of QCD species 
produced via Hawking radiation. Interactions among emitted particles 
are not able to produce a thermal 
shell of radiation, so the spectrum of fundamental species exit the
BH with basically the black body spectrum described by Hawking. 
The final spectrum of stable particles at large distances, however,  
is peaked around $\Lambda_{QCD}$, and exhibits features weakly 
dependent upon number of extra dimensions or the BHs mass. 
Standard Model 
modes are constituted by an approximate $43\%$ of neutrinos, 
a $28\%$ of electrons, a $16\%$ of photons and a $13\%$ of protons. 
The gravitons produced are a fraction that goes from the $0.4\%$ of the 
total number of particles ($16\%$ of the energy) for $M=10^{10}$ GeV 
and $n = 6$ to the $0.02\%$ ($1.4\%$ of the energy) for $M=10$ TeV and $n=2$.

This work is a preliminary analysis, with results that can be 
useful for future search for effects of TeV gravity on cosmic ray
physics.

\end{document}